\documentclass{PoS}

\usepackage{url}
\usepackage{amsmath}
\usepackage{amssymb}
\usepackage{graphicx}
\usepackage{bm}
\usepackage{epsfig}
\usepackage{amsfonts}
\usepackage{color}
\usepackage{mathtools}
\usepackage[section]{placeins}
\usepackage{wrapfig}
 \usepackage{notoccite}
 \usepackage[numbers]{natbib}
 
\title{Bayesian techniques and applications to QCD}

\ShortTitle{Bayesian techniques and applications to QCD}

\author{\speaker{Alexander Rothkopf}\\
        Faculty of Science and Technology, University of Stavanger, NO-4036 Stavanger, Norway\\
        E-mail: \email{alexander.rothkopf@uis.no}}

\abstract{Realizing the full potential of interconnecting the large amounts of data created in physics experiments, phenomenological models and theory simulations requires robust tools for statistical inference. Here I review a particularly promising branch, Bayesian statistics, which over the past decade has found manifold use in high-energy physics. After a brief introduction to Bayesian statistics I will present two concrete examples, where Bayesian thinking has led to progress in understanding strongly interacting matter: unfolding problems in the form of lattice QCD spectral functions (in spirit similar to detector corrections), as well as the efficient estimation of quark-gluon-plasma parameters from a systematic comparison of experimental heavy-ion collision data and phenomenological models.}

\FullConference{XIII Quark Confinement and the Hadron Spectrum - Confinement2018\\
		31 July - 6 August 2018\\
		Maynooth University, Ireland}

\begin{document}

\section{Motivation}

With the advent of modern high energy collider facilities, be it RHIC at BNL or the LHC at CERN, high energy physics is able to tap into vast amounts of empirical data. On the other hand high-performance computing has made possible the computation and prediction of many phenomenologically relevant quantities from first principles (e.g. hadron masses \cite{Borsanyi:2014jba} and thermodynamics \cite{Bazavov:2017dus,Borsanyi:2016ksw}) . Concurrently phenomenology has progressed in constructing more and more refined models, take for example the treatment of the dynamical evolution of a heavy-ion collision (HIC), which allows us to reproduce many features observed in experiment (for an introduction see e.g. \cite{Florkowski:2014yza}). Making these three treasure troves mutually accessible using efficient methods of statistical inference, in part supported by rapidly developing technologies such as machine learning, is a central goal for theoretical physics in the 21st century. As one of the first high-energy theory conferences, ConfXIII hosts a full session\footnote{\href{https://indico.cern.ch/event/648004/sessions/266239/#all}{ConfXIII - Statistical Methods for Physics Analysis in the XXI Century session}} dedicated to this topic.

Bayesian statistics and inference based upon it is one particularly useful piece in the arsenal available to the avid physicist in approaching the challenge of distilling novel physics insight from empirical and simulation data. For those interested in joining this field \cite{gelmanbda04,mcelreath2016statistical} can serve as a first starting point.

The intention of this talk is to contribute towards demystifying the concept of Bayesian statistics and to provide two concrete examples, where it has helped improve our understanding of strongly interacting matter and QCD.

\section{Bayesian inference} \vspace{-0.75cm}
\begin{figure}
\centering
 \includegraphics[scale=0.5, trim= 0cm 11.5cm 11cm 0.6cm, clip=true]{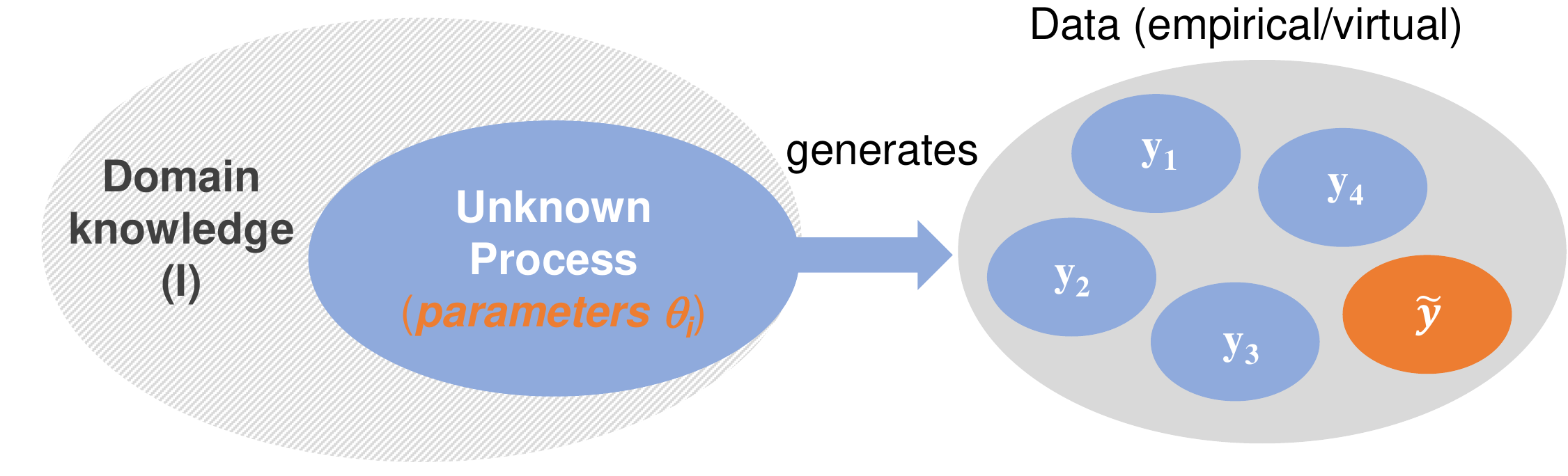}
 \caption{Generic setup for statistical inference. Bayesian inference explicitly acknowledges that a model of the unknown process is embedded in a domain to which prior knowledge $I$ exists.}\label{Fig:Scenario}
\end{figure}
Let me start with a very general scenario (see Fig.\ref{Fig:Scenario}), where we are observing some unknown process generating data $y_k$. This may be an experimental setup or a simulation based on Monte-Carlo sampling. The central goal of \textit{statistical inference} now is to draw conclusions about unobserved quantities, based on that empirical data. What we mean by unobserved quantities are on the one hand the unknown parameters of the generating process $\theta_i$ and on the other hand future, i.e. as of yet unobserved data $\tilde y_l$.

\textit{Bayesian inference} is a particular form of statistical inference, which uses a generalized concept of probability. Probability in the Bayesian sense is understood as a measure of uncertainty and does not need to be directly related to a large number of repeatable trials. The flexibility of the Bayesian approach may be glimpsed when considering the question: What is the chance of the first launch of the new Ariane 6 rocket to succeed. Bayesian statistics sets out to answer in the form of a probability statement. At first it may seem an improper task, as we cannot base our answer on a large number of trials of first launches (not to mention the economic difficulty).

However whenever we describe a process in nature, we are forced to model it using the language of physics and mathematics. Such a model in turn exists within the context of the process of interest, in this case aerospace engineering. And of course there exists prior \textit{domain knowledge} $I$ that informs our model. E.g. engineers have tested individual components of the Ariane rocket, such as valves, for their reliability using e.g. frequentist approaches. As we shall see, Bayesian statistics asks us to explicitly state what the uncertainties of or model parameters are and helps us propagate these uncertainties into the final answer in the language of a generalized probability.

The starting point for Bayesian inference is the joint probability distribution of the involved quantities $p(\theta,y,I)$. Constructing it can be a formidable task, as it requires knowledge about the data generating process and the physics context it is embedded in. Using the rules for conditional probabilities we may reexpress $p(\theta,y,I)$ to give way to \textit{Bayes theorem}
\begin{align}
 \underbracket{p(\theta|y,I)}_{{\rm posterior}}=\underbracket{p(y|\theta,I)}_{{\rm likelihood}}\underbracket{p(\theta|I)}_{{\rm prior}}/\underbracket{p(y|I)}_{{\rm evidence}},
\end{align}
whose parts are straightforwardly matched to our problem at hand. The likelihood encodes how probable it is that a datapoint $y$ has been generated from a process with fixed set of parameters $\theta$. The prior probability on the other hand tells us how probable it is that a certain set of $\theta$ parameters arises from our prior domain knowledge. The evidence (its name is simply a historic convention) enters as a normalization factor independent of the parameters $\theta$.

To the frequentist scholar the prior probability might be a stumbling block, so let us have a closer look. In the literature $p(\theta|I)$ is often parameterized using two quantities, the \textit{default model} $m$ and a weight $\alpha$. The former refers to the extremum of the prior $\partial_\theta p(\theta|I)|_{\theta=m} =0$ while $\alpha$ denotes an overall weight factor $p(\theta|I)={\rm exp}(\alpha P(\theta|I))$, i.e. they take the role of a \textit{hyperparameter}. Bayesian statistics now forces us to make explicit the parameters of the prior in the joint probability distribution $p(\theta,y,m,\alpha)$. In modern approaches to Bayesian statistics one acknowledges that model parameters themselves carry uncertainties, so that we must endow both $m$ and $\alpha$ with distributions of their own, so called hyperpriors $p(m)$ and $p(\alpha)$. This is the simplest form of a so called \textit{hierarchical model} in which successive layers inform us about the interdependencies and uncertainties of the parameters entering our model.

Once the prior and posterior have been determined we are able to close in on the central task of Bayesian inference, to compute (or more likely numerically estimate) the posterior distribution, which provides access to
\begin{align}
 p_{\rm MP}(\theta_j|y,I)=\Pi_{i\neq j}\int d\theta_i p(\theta|y,I), \quad p_{\rm PP}(\tilde y| y,I)=\int d\theta p(\tilde y| \theta,I)p(\theta|y,I),
\end{align}
the \textit{marginal posterior} $p_{\rm MP}(\theta_j|y,I)$ and the \textit{posterior predictive} distribution $p_{\rm PP}(\tilde y| y,I)$. The former constitutes the final say of Bayesian inference on how probable a certain parameter of our unknown process is given the observed data and prior knowledge. The latter is valuable for assessing the validity of the inference, as its predictions for future measurements can be compared against actual collected data.

At this point I would like to stress that in its modern incarnation Bayesian inference does not suffer from what is often called subjectivity in the literature. We have seen that prior knowledge is nothing more than domain knowledge with the appropriate uncertainties encoded via hyperpriors. 

What makes the Bayesian approach very efficient is the fact that once new measurements $y^\prime$ become available they are easily incorporated in an updated analysis using the old posterior as the new prior $p(\theta|y^\prime,y,I)\propto p(y^\prime|\theta,I)p(\theta|y,I)$. Obviously knowledge of the full posterior probability $p(\theta|y,I)$ is needed in this case.

The past decades have seen major progress in the application of Bayesian statistics, often due to improved computational capabilities. Among them is the consistent adaption of priors arising from genuine domain knowledge and not from to computational convenience (i.e.\ a liberation from conjugate priors). The use of multi-layered hierarchical models nowadays allows us to self-consistently incorporate the full inter dependencies and uncertainties of model parameters. Last but not least, the availability of powerful numerical libraries based on Monte-Carlo techniques has made it possible to straight forwardly sample the full posterior distribution instead of settling for just determining its extremum (maximum aposteriori, MAP) as was often the case in the older literature.

Indeed a full Bayesian analysis today is only one download away with e.g.\ the \href{http://mc-stan.org/}{MC-STAN} library\footnote{Thanks go out to N. Wink from Heidelberg for bringing this library to my attention} \cite{JSSv076i01}. This software, maintained and supported both by scientific organizations, such as the Max-Planck Society, as well as industry in the form of Google, provides a high level language to formulate hierarchical models and to evaluate their posterior via an efficient Hamiltonian (hybrid) Monte Carlo sampler (No-UTurn). 

After having briefly gone through the main ingredients to Bayesian inference let us continue by considering two concrete examples where Bayesian thinking has proven insightful in the context of QCD.

\section{Applications to QCD}

\subsection{Lattice QCD spectral function unfolding}

The topic discussed in this subsection falls into the category of inverse problems that come in the form of an unfolding task. Extracting from the readings of a detector the actual trajectory of a charged particle (see related contributions \cite{Kruger,Kuusela}), as well as the extraction of spectral functions in lattice QCD simulations (see related contributions \cite{Kaczmarek,Astrakhantsev,Quinn,Rothkopf}) are but two examples.

The starting point is to consider how our data is obtained, i.e.\ how to formulate the likelihood. In both cases the observed data $\vec y$ (track in detector, simulated correlation function) is related to the actual quantity of interest $\vec{x}$ (actual track of particle, spectral function) via a linear transformation $\vec{y}=\hat K \vec x +\vec\eta$. The kernel $K$ represents the imperfections of the detector or is a manifestation of Euclidean quantum field theory. It may be obtained from e.g.\ detector simulations or in the case of lattice QCD correlators from QCD directly. In both experiment and Monte-Carlo simulations the data carry a finite precision, which may be captured by an additive noise term $\vec\eta$. In a lattice QCD simulation it often turns out to be Gaussian distributed, which leads to a likelihood that itself is of Gaussian form 
\begin{align}
 p(y|x)\propto {\rm exp}\big[(y-Kx)^2/2\sigma^2\big].
\end{align}
Here $\sigma$ denotes the uncertainty in the input data $y$. Of course in a real-world detector setup the type of noise may differ significantly.

If we were to attempt a simple $\chi^2$ fit of $\vec x$ from the data $\vec y$, it would correspond in the language of Bayes to setting the prior probability to unity and determining the maximum of the corresponding posterior. This is what is known as a \textit{maximum likelihood} estimate. The difficulty of inverse problems lies in the fact that they are often ill-posed, so that the maximum of the likelihood is not unique and further regularization of the problem is required.

\begin{figure}
\centering
 \includegraphics[scale=0.5, trim= 0cm 11.5cm 14.5cm 0.6cm, clip=true]{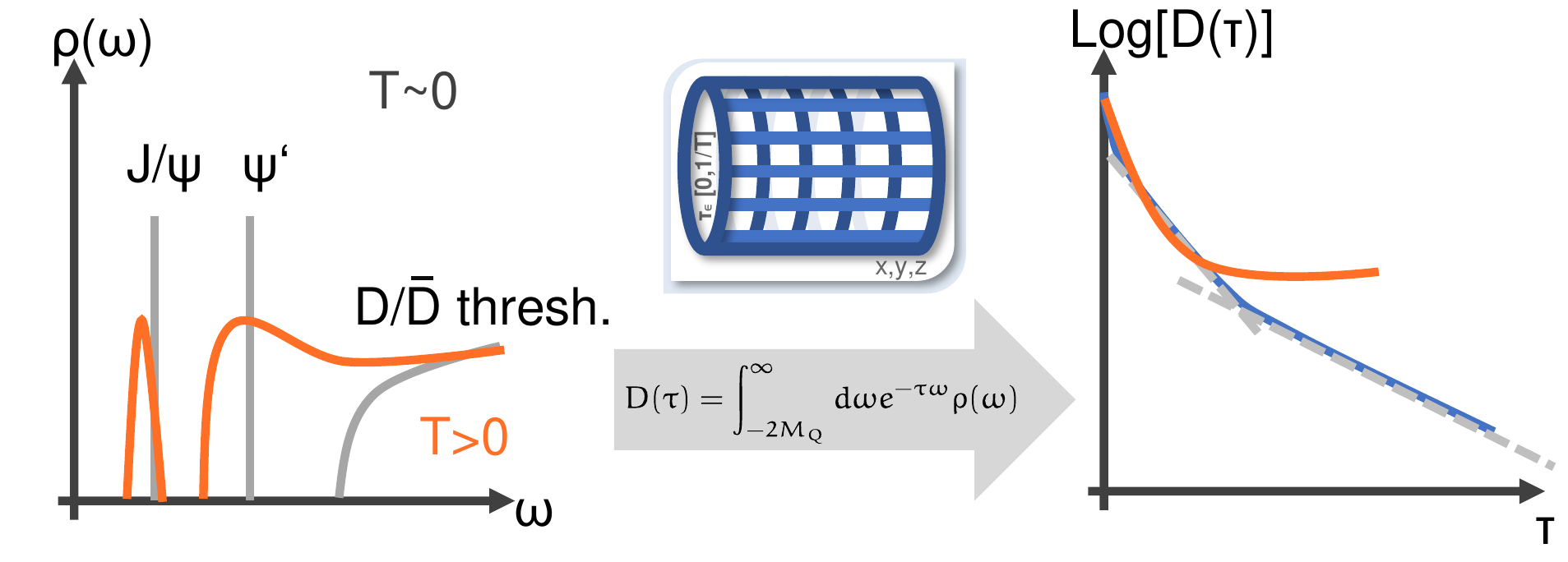}
 \caption{Translating a physical hadron spectral function into a Euclidean correlator via lattice QCD.}
 \label{Fig:lattice}
\end{figure}

Take the task of reconstructing spectral functions from lattice QCD simulations as concrete example \cite{Jarrell:1996rrw}. Numerical simulations of strongly interacting matter on the lattice act very similar to a very distorting detector setup. On the left in Fig.\ref{Fig:lattice} I have sketched a hadronic spectrum in vacuum in gray, where we see two well defined bound states and a continuum structure at higher frequencies. In a lattice QCD simulation, which is carried out in an artificial imaginary time direction, this spectrum is translated into a Euclidean correlation function, plotted on the right in blue.
The kernel in this case (the example pertains to NRQCD \cite{Lepage:1992tx} fermions) amounts to a Laplace transformation and thus is exponentially damped with frequency $\omega$ and Euclidean time $\tau$. In vacuum and in the presence of well separated peaks the bound states manifest themselves as well pronounced exponential falloffs, indicated by the gray dashed lines on the right. This may allow an exponential fitting procedure to reveal many of the vital properties encoded in such a correlator. 

Once we move to finite temperature, which is where we wish to learn about QCD for heavy-ion collisions, the spectrum will change significantly. In-medium bound states will show broad peaks, some will already have melted or persist as a mere threshold enhancement of a much more dominant continuum (solid orange line). This translates into a correlator where no more clear exponential behavior is visible and which instead shows overall curvature. The most limiting aspect of the Euclidean formulation of thermal field theory is that temperature and the length of the accessible Euclidean time axis are inversely proportional, indicated by the orange correlator only reaching to a much smaller maximum $\tau$ value.

The task at hand is to extract from the simulated correlators the spectral function, containing the relevant physical information about hadrons at finite temperature. This inverse Laplace transform is \textit{the} classic ill-posed inverse problem, in particular, since the intricate structures present in the spectrum require us to discretize frequencies with many more bins $N_\omega$ than we have simulation data points $N_\tau$.
\begin{align}
 D(\tau_i)=D_i=\sum_{l=1}^{N_\omega} \, \Delta\omega_l {\rm exp}[-\omega_l \tau_i] \, \rho_l.\quad N_\omega\gg N_\tau,\quad D\sim N(\hat K \vec \rho,C) \label{Eq:DiscSpecRep}
\end{align}
Thanks to the law of large numbers the correlator data $D_i$ are distributed according to a (correlated) Gaussian distribution.

In order to provide a regularization in the Bayesian sense, we need to specify a prior probability which will yield a well behaved, i.e. unimodal posterior. There are several proposals on the market on how to construct such a prior. One proposal arises from what otherwise is known as TV regularization \cite{RUDIN1992259}, which in the language of Bayes can be recast as $p_{\rm TV}(\rho|m,\alpha)={\rm Laplace}[m,1/\alpha]$. Standard Tikhonov regularization \cite{MacKay:2002} on the other hand is equivalent to $p_{\rm Tk}(\rho|m,\alpha)=N[0,1/\alpha]$ and may be straight forwardly modified to actually take into account the values of the default model $p_{\rm Tk'}(\rho|m,\alpha)=N[m,1/\alpha]$. The statement that standard Tikhonov regularization does not introduce a default model is incorrect, as it simply sets the values of the default model to zero, leading to the phenomenon of parameter shrinkage.

The most popular choice to date is the Maximum Entropy Method (MEM) \cite{Asakawa:2000tr} with \linebreak $p_{\rm MEM}(\rho|m,\alpha)\propto e^{S_{\rm SJ}}$, which employs the Shannon-Jaynes entropy $S_{\rm SJ}(m,\alpha)=\alpha\int\big(\rho-m$ \linebreak $-\rho{\rm log}[\rho/m]\big)d\omega$. This prior is based on four axioms motivated from considerations of image reconstruction. Contrary to the previous two proposals it actually enforces positivity, which is a genuine piece of prior information in case of hadronic spectra in QCD. It has been argued that the principle of maximum entropy which motivates $S_{\rm SJ}$ leads this prior to favor smooth spectral functions, an assertion which has however been proven incorrect \cite{Kim:2018yhk,Fischer:2017kbq}.

A recent addition to this collection of priors is the BR prior \cite{Burnier:2013nla} $p_{\rm BR}(\rho|m,\alpha)={\rm Gamma}[\alpha+1,\alpha/m]$ (abbreviated from Bayesian Reconstruction), which is also based on four axioms. In their motivation however the one-dimensional reconstruction task takes center stage. Again positivity is a vital ingredient. In addition the requirement of scale invariance plays a central role, which takes care of the fact that a spectral function (contrary to statements in the literature) is not in general a probability distribution. (In particular hadronic spectral functions diverge with positive powers of the frequency, as shown in perturbation theory.) From all the proposals so far the BR prior imprints prior information in the most gentle fashion, while still leading to a unimodal posterior. A generalization of the BR prior to non-positive spectral functions \cite{Rothkopf:2016luz}, encountered e.g. in the study of parton spectra \cite{Cyrol:2018xeq,Ilgenfritz:2017kkp} has also been put forward.

Once the prior is specified the standard procedure in the literature up to now is to compute the maximum of the posterior (MAP) and declare this function the Bayesian answer to the most probable spectrum, given simulation data and prior information. As long as the posterior is not heavy tailed this is a valid strategy. (For the technical details on how the hyperparameter $\alpha$ is treated in the different methods see \cite{Asakawa:2000tr,Burnier:2013nla}). It turns out that the MAP solution may suffer from ringing artifacts as long as local regulators, such as those in the MEM and BR method \cite{Kim:2018yhk} are used. The exploration of non-local regulators (see e.g. \cite{Kim:2018yhk,Fischer:2017kbq}), which in the language of Bayesian statistics correspond to correlated distributions among the different parameters $\rho_l$ is work in progress.

Nowadays we are however not limited to the simple MAP estimate and can easily explore the full posterior distribution using e.g. the MC-STAN framework. This also liberates us from some of the rather ad-hoc approximations deployed when handling the hyperparameter $\alpha$ e.g. in the MEM.\begin{wrapfigure}{r}{0.32\textwidth}\vspace{-1cm}
  \begin{center}
   \includegraphics[scale=0.35]{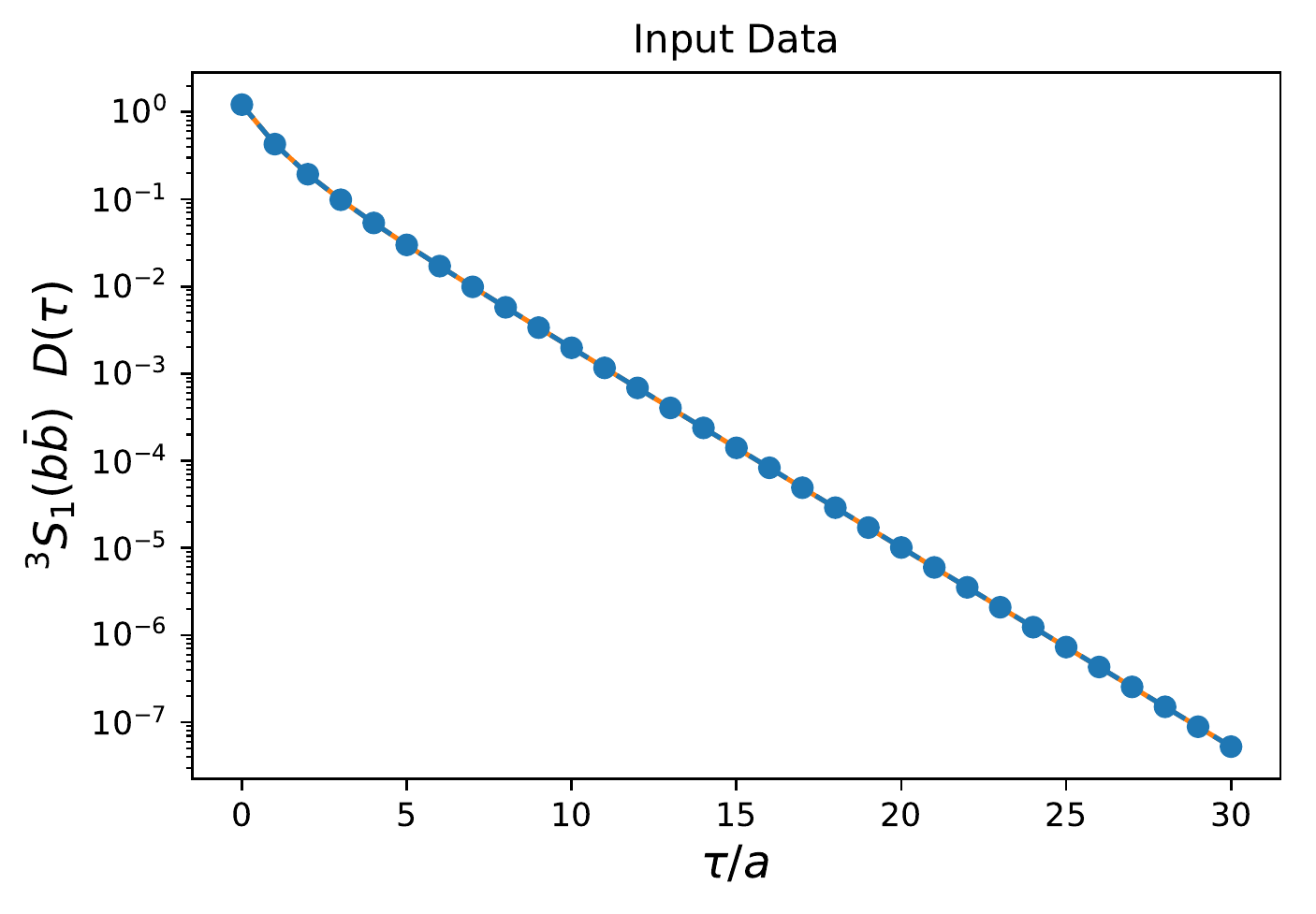}
   \includegraphics[scale=0.35]{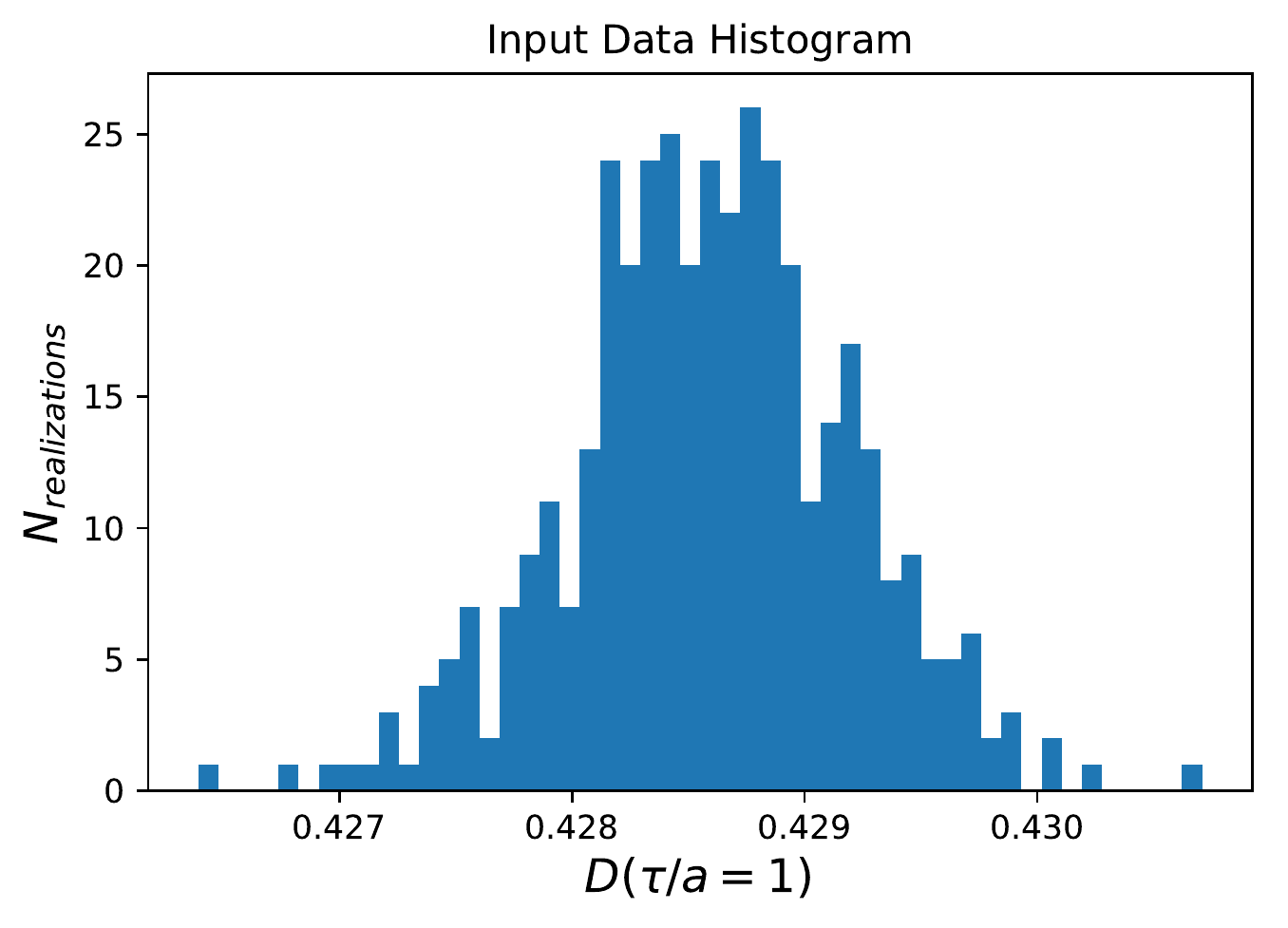}
   \includegraphics[scale=0.35]{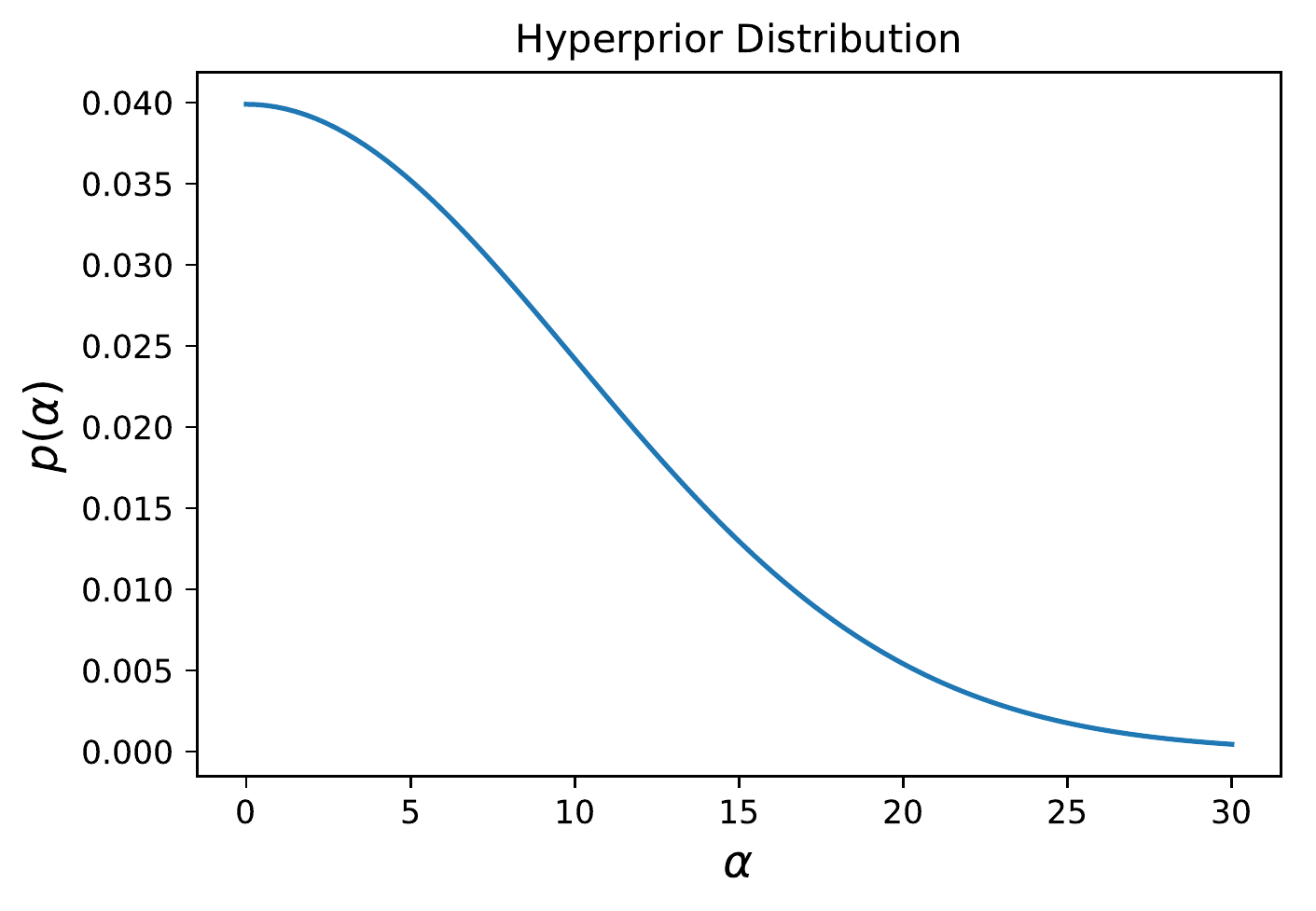}
   \includegraphics[scale=0.35]{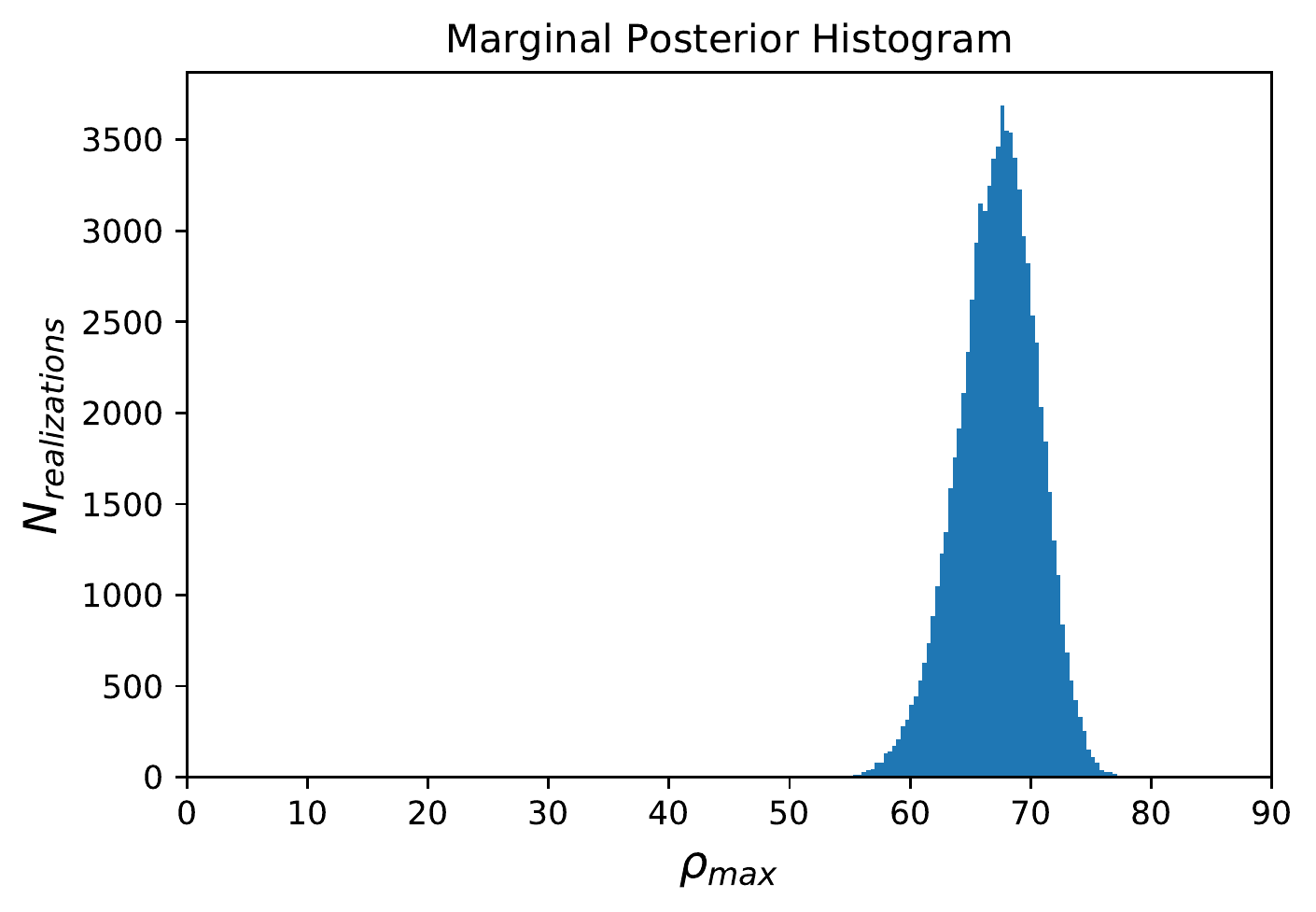}
   \includegraphics[scale=0.35]{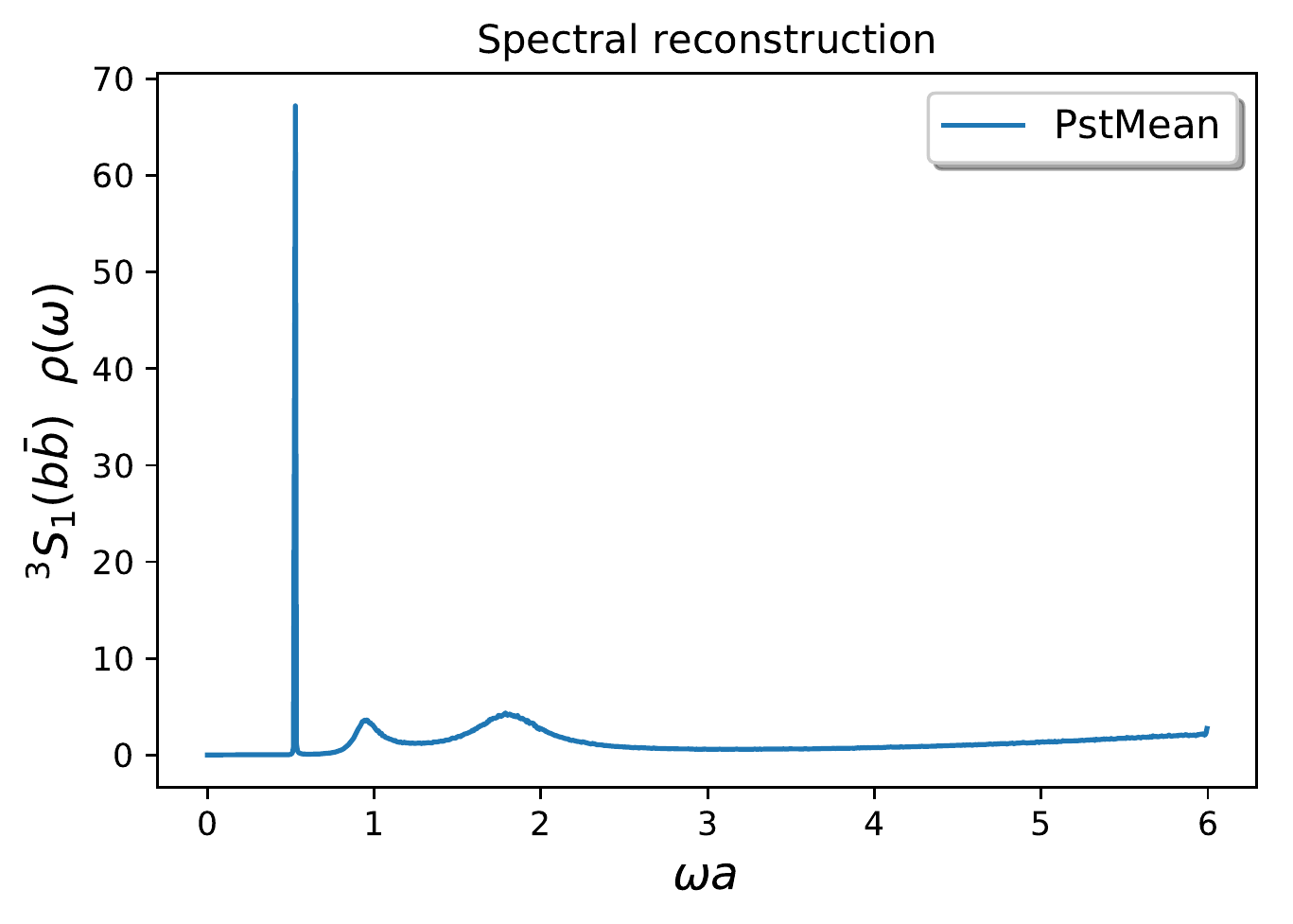}
   \includegraphics[scale=0.35]{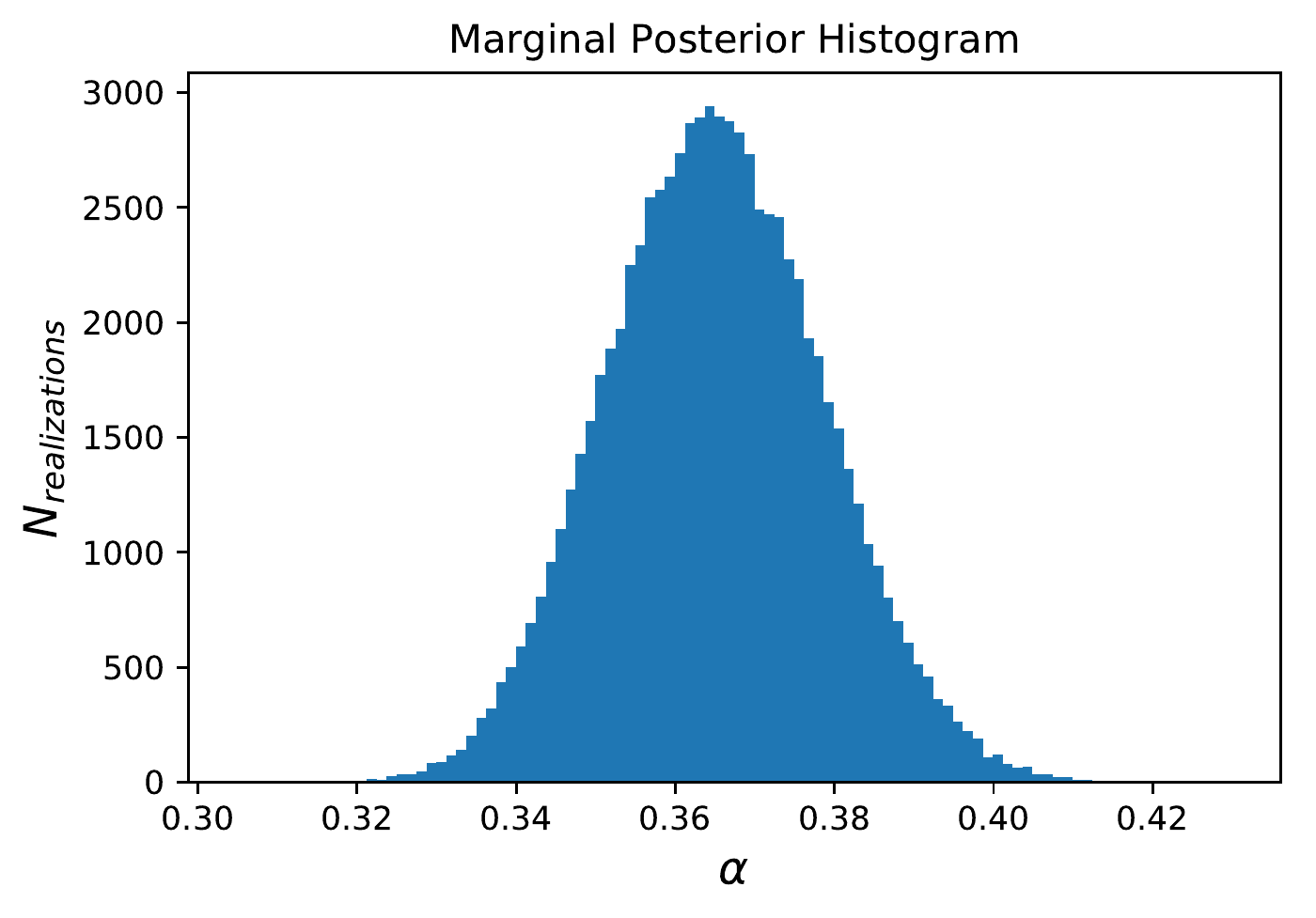}
  \end{center}\vspace{-0.9cm}
 \caption{Relevant parts of the Bayesian analysis: input data mean, histogram of $D_1$, $p(\alpha)$, $p(\rho_{\rm max}|D,I)$, $\rho_{\rm Bayes}$, $p(\alpha|D)$.}\label{Fig:Bayes}\vspace{-10.cm}
\end{wrapfigure} Let me give a concrete example here using actual lattice NRQCD correlation functions for the Bottomonium $^3S_1$ state Upsilon (data from \cite{Kim:2018yhk}) to carry out a Bayesian reconstruction with the BR prior.

We start by computing the $N_\tau$ means $D_i$ (see top panel of Fig.\ref{Fig:Bayes}) of the $N_{\rm data}$ correlator samples $D^k_i$ (see 2nd panel in Fig.\ref{Fig:Bayes} for $D_1^k$) and their correlation matrix $C_{ij}=\frac{1}{N_{\rm data}(N_{\rm data}-1)}\sum_k(D^k_i-D_i)(D^k_j-D_j)$ (note that this is the correlation matrix of the mean with respect to the correct value $D^{\rm true}_i$). The relation between this data and the discretized spectral function is given by eq.\eqref{Eq:DiscSpecRep}. For computational efficiency we can apply a coordinate transformation on the kernel matrix $\hat K$ and the data vector $\vec D$ to a basis in which the correlation matrix $C$ is diagonal. Then we may consider each $\tilde D_i$ as truly independently Gaussian distributed with an uncertainty given by the corresponding eigenvalue of $C$. This sets the likelihood of the problem.

Now we turn to the BR prior, to which we here for simplicity supply a constant default model $m=1$. In practice one can incorporate into $m$ e.g. the large $\omega$ behavior of the spectrum known from lattice perturbation theory. The prior includes a hyperparameter $\alpha$, whose values we do not know apriori. This uncertainty we make explicit by using a very broad half Gaussian distribution as hyperprior $p(\alpha)={\rm HalfNormal}[0,10]$ (3rd panel of Fig.\ref{Fig:Bayes}). For $\alpha=10$ the prior should already fully dominate the posterior. This is all that is necessary to set up the MC-STAN model which reads
\begin{verbatim}
 stan_model_BR = """
data {
    int sNt; // # of tau input data
    int sNw;  // # of rho bins
    matrix[sNt, sNw] Kernel;
    vector[sNt] D;  // simulation data
    real<lower=0> ConstDM; // default model
    vector[sNt] Uncertainty;  // corr. error
}
parameters {
    vector<lower=0>[sNw] rho;
    real<lower=0> alpha; // Hyperparameter
}
model {
    alpha ~ normal(0, 10); //hyperprior
    rho ~ gamma(alpha+1, alpha/ConstDM);
    D ~ normal(Kernel * rho, Uncertainty);
}
"""
\end{verbatim}
These lines are all that is needed to specify a full Bayesian analysis. Using $N_\omega=1000$ frequency bins with $N_\tau=32$ and running with 40 MC chains of 2500 iterations each (500 warmup steps) we obtain 
a well defined histogram for each of the marginal posteriors $p(\rho_l|D,I)$ (4th panel in Fig.\ref{Fig:Bayes} for the maximum $\rho_l$). Via their mean a highly accurate image of the encoded spectral function emerges (panel 2nd from bottom in Fig.\ref{Fig:Bayes}). Note the dominant peak, which corresponds to the exponential falloff in the correlator.

Since we here treated $\alpha$ as one of the parameters of the joint probability distribution we may actually learn self-consistently, how strongly the regulator affects the posterior by looking at the marginal posterior $p(\alpha|D)$ shown in the bottom panel of Fig.\ref{Fig:Bayes}.

Even if the implementation of a Bayesian analysis nowadays is straight forward several conceptual challenges remain. One is related to the question of the information content within the lattice QCD correlation functions. Especially at  $T>0$ the finite Euclidean time extend limits the accuracy of the reconstruction and going to the continuum limit does not remedy this issue (for a recent discussion see \cite{Kim:2018yhk,Pawlowski:2016eck}). The question to answer is how to set up simulations to improve the relevant information content. Possible strategies involve going to anisotropic lattices or more exploratory routes, such as attempting to generalize the multi-level algorithm to full QCD \cite{Ce:2016ajy} or to explore modified time contours in the complex plane \cite{Pawlowski:2016eck}.

The other challenge is how to incorporate analytic properties of the spectra and correlators into the regulator. Current priors use mostly concepts unspecific to QCD, e.g. smoothness. On the other hand analytic properties may be treated by introducing e.g. restricted sets of basis functions among which to select the MAP \cite{Cyrol:2018xeq}. This however is not Bayesian in nature, where prior information must be provided in terms of a functional or correspondingly a probability distribution. One possible strategy is to deploy machine learning to construct an appropriate regulator from prior knowledge, which is work in progress.

\subsection{Quark-gluon-plasma parameter estimation}

The second example of Bayesian inference in QCD connects the progress made in the experimental investigation of heavy-ion-collisions with that of phenomenologically modelling the dynamical evolution of the bulk matter created therein. Its physics goal is to generate new insight into the properties of the quark-gluon-plasma. 

Over the past decade a \textit{standard model} of HICs has emerged whose success rests on chaining together several effective descriptions of the physics relevant at different stages of the collision. It starts from fluctuating initial conditions, modern incarnations of which are based on classical statistical simulations of Yang-Mills fields. These smoothly connect via kinetic theory to a description of the bulk matter in terms of viscous relativistic hydrodynamics that in turn terminates at hadronization, where hadronization models connect to transport models for the hadronic phase. At the same time highly precise measurements of the yields and correlations among light hadrons (e.g. harmonic flow) have been obtained in experiment. 

It is natural to ask how statistical inference can help us identify the most probable values for the parameters entering the phenomenological models of HICs by employing Bayesian reasoning. We here follow the strategy laid out in the first \cite{Bernhard:2016tnd} in a series of studies \cite{Auvinen:2017fjw} approaching this question in a genuinely Bayesian fashion (see related contribution \cite{Kovalenko}). The problem at hand is well suited for Bayesian statistics as we have to deal with many sources of uncertainties, stemming in part from finite precision in theory input, such as e.g. in the equation of state of QCD obtained from lattice QCD. On the other hand one is faced with e.g. parametrization choices for quantities, such as the specific viscosities, since no theory consensus has been reached on their temperature dependence so far. The one uncertainty that is not part of such an analysis is the one related to the range of validity of the deployed models themselves, i.e. it cannot be quantified in how far e.g. viscous hydrodynamics is an appropriate description of the intermediate time evolution of the QGP.

The authors of \cite{Bernhard:2016tnd} choose nine parameters among all model parameters which are left free and fix all others using best estimates from the literature. While strictly speaking this does not amount to a fully Bayesian approach it is a very important first step into the direction of an urgently needed comprehensive Bayesian parameter estimation. The free parameters come in two groups, one related to the initial conditions containing four quantities: $p$ entropy deposition, $k$ shape fluctuations, $w$ Gaussian nucleon width and a normalization. The other five parameters are genuine QGP properties: $T_{\rm switch}$ hadronization temperature, $\eta/S_{\rm HRG}$ specific viscosity in the hadronic phase, $\eta/S_{\rm min}$ minimal specific viscosity at $T_{\rm switch}$, the slope of the specific viscosity modeled as linear increasing function of T and $\zeta/S$ the magnitude of the specific bulk viscosity.

Given these 9 input parameters $\theta_i$ the chain of models is able to produce a plethora of simulated data $y_{\rm sim}$, such as yields $dN_\pi/dy$ or $v_2$ for different centrality bins which may then be compared to experimental measurements. Bayes theorem in this case then takes the form 
\begin{align}
p(\theta|y_{\rm exp})\propto p(y_{\rm exp}|\theta)p(\theta),
\end{align}
where we have suppressed the dependence of the likelihood and posterior on the choice of models used.

The main focus in the following lies on how to efficiently describe the likelihood $p(y_{\rm exp}|\theta)$ in this problem for which \cite{Bernhard:2016tnd} deploys a technique that is well established in the realm of machine learning: Gaussian processes. An introduction to the general topic can be found in \cite{rasmussen2006gaussian}.

The challenge lies in the fact that to exhaustively sample the posterior we need to evaluate the likelihood for a very large range of parameter values. At the same time the computational cost involved in running the chain of models prevents us from doing so in any reasonable time frame. In order to overcome this limitation, Gaussian processes provide a strategy of how to efficiently estimate the true likelihood based on a small number of simulated training data sets. The starting point is to regard the simulated data $y_{\rm sim}$ as distributed according to a correlated Gaussian, where the non-trivial information is contained in a usually dense correlation matrix $\Sigma$. The corresponding likelihood then also takes the form of a correlated Gaussian
\begin{align}
p(y_{\rm exp}|\theta)\approx {\rm exp}\big[ - (y_{\rm sim}(\theta)-y_{\rm exp})\Sigma_{\rm GP}^{-1} (y_{\rm sim}(\theta)-y_{\rm exp}) \big].\label{Eq:ApproxLikelihood}
\end{align}
As preparatory step one carries out full model computations on a set of ${\cal O}(100)$ model parameter sets $\theta_{\rm train}$ distributed for efficiency on a \textit{Latin hypercube}. These will serve as training data $y_{\rm train}$ for the Gaussian process.

Our goal is to set up an approximate Gaussian distribution for $y_{\rm sim}(\theta)$ which is fully specified by its mean $m(\theta,\theta_{\rm train})$ and its correlations $K(\theta,\theta_{\rm train})$ interpolating between the values $\theta_{\rm train}$. The starting point is to assume that the training data belongs to the same Gaussian distribution from which we will draw our estimates $y_{\rm sim}(\theta)$ for some other parameter set  
\begin{align}
 \left( \begin{array}{c} y_{\rm train}\\ y_{\rm sim}\end{array}\right)=N\left(\begin{array}{c} m(\theta)\\ m(\theta,\theta')\end{array}, \begin{array}{cc} K(\theta_{\rm train},\theta_{\rm train})& K(\theta_{\rm train},\theta_{\rm sim})\\ K(\theta_{\rm sim},\theta_{\rm train})& K(\theta_{\rm sim},\theta_{\rm sim}) \end{array}\right).
\end{align}
All the important information resides in the model correlation functions $K$, which we here take to also have Gaussian form. This is a particular choice and independent from the Gaussian character of the joint probability distribution
\begin{align}
 K(\theta_k,\theta_l')=\sigma_{\rm GP}^2{\rm exp}\big[-\frac{(\theta_k-\theta_l')^2}{2l^2}\big]+\sigma_n^2\delta_{kl}.
\end{align}
We have three hyperparameters in this model, $\sigma_{\rm GP}$ specifying the strength of correlation among the training data, the correlation lengths $l_k$ characterizing the distance between parameters and $\sigma_n$ related to the uncertainty of the $y_{\rm train}$'s. It is the values for these, which we need to fit using the training data.

\begin{wrapfigure}{r}{0.32\textwidth}\vspace{-1.4cm}
\begin{center}
 \includegraphics[scale=0.7, trim= 0cm 0cm 11.8cm 0cm, clip=true ]{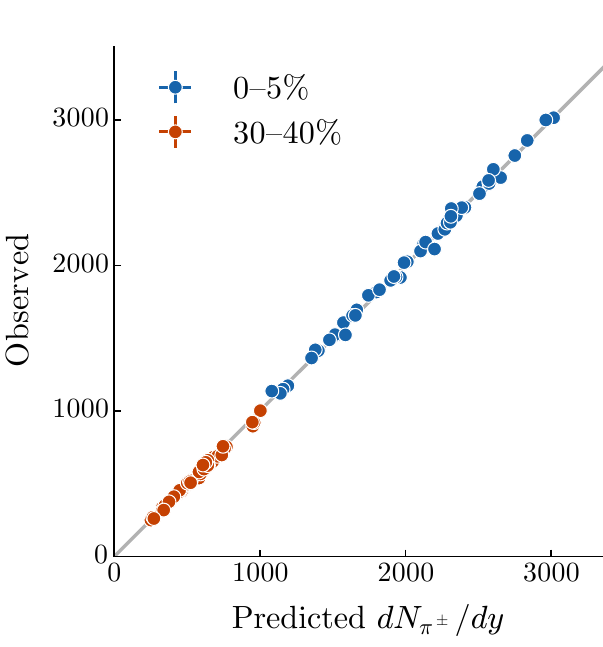}\vspace{0.2cm}
 \includegraphics[scale=0.55]{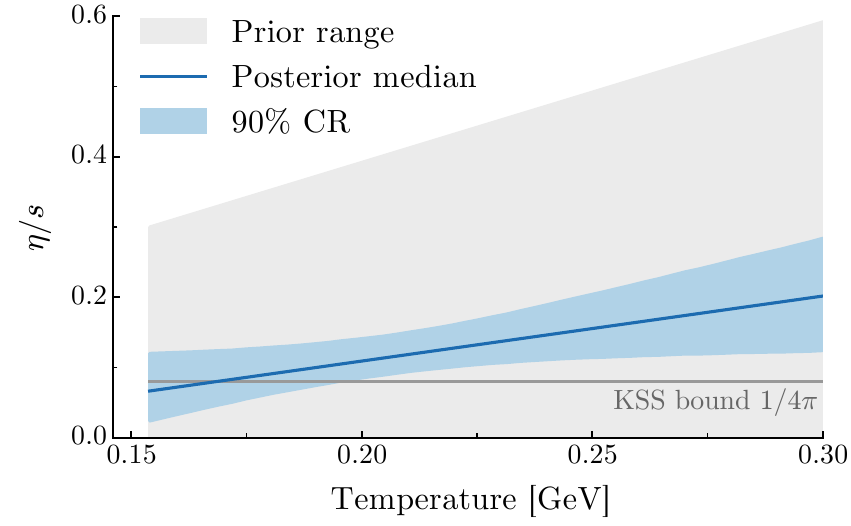}
  \end{center}\vspace{-0.6cm}
 \caption{(top) Validation of the Gaussian process model using experimental data (bottom) specific shear viscosity estimate from the posterior (figures taken from \cite{Bernhard:2016tnd}) }\label{Fig:QGPAnalysis}\vspace{-0.6cm}
\end{wrapfigure}Using $p(z|x) = p(x,z)/\int dz p(x,z)$ we then arrive at the conditional probability for our estimated $y_{\rm sim}$ as $p(y_{\rm sim}(\theta)|y_{\rm train})=N(\mu_{\rm GP}(\theta),\Sigma_{\rm GP}(\theta))$ with expressions for 
\begin{align}
 \nonumber&\Sigma_{\rm GP}=K(\theta,\theta)-K(\theta,\theta_{\rm train})-K(\theta_{\rm train},\theta_{\rm train})^{-1}K(\theta_{\rm train},\theta_{\rm train}),\\
 \nonumber &\mu_{\rm GP}= m(\theta_{\rm train}) + K(\theta,\theta_{\rm train})K(\theta_{\rm train},\theta_{\rm train})^{-1} (y_{\rm train}-m(\theta_{\rm train})).
\end{align}
I.e. we now are in possession of a means to quickly draw an approximate $y_{\rm sim}$ to be used in eq.\eqref{Eq:ApproxLikelihood} together with the appropriate correlation function $\Sigma_{\rm GP}$.

Before we deploy the Gaussian process let us make sure that the approximation made so far is not too simplistic and manages to capture the relevant physics encoded in $y_{\rm train}$. To this end one can evaluate the full model computations at a new set of parameters and compare those to the predictions. As shown in the top panel of Fig.\ref{Fig:QGPAnalysis} the correlation between prediction and actual values is excellent, the values scatter tightly along the diagonal.

The last missing piece to the Bayesian analysis is to specify the prior distribution. Here the authors resort to uniform distributions for the parameter values setting the upper and lower limits large enough so that the relatively large uncertainty e.g. in the specific viscosities (gray shaded region in Fig.\ref{Fig:QGPAnalysis} bottom panel) is well represented. Combining these together with the likelihood we constructed before allows us to systematically interrogate the experimental data and infer the most probable value of the parameters (blue line in bottom panel of Fig.\ref{Fig:QGPAnalysis}) accompanied by a significantly reduced uncertainty band. 

\section{Conclusion}

In this contribution my goal has been to argue that Bayesian inference provides a flexible approach to extract insight from empirical data based on the consistent application of Bayes theorem. In the introductory part I showed that in its modern incarnation Bayesian statistics does not suffer from subjectivity, as prior information is understood as domain knowledge informing our modelling with the appropriate uncertainties properly encoded in hyperparameters of hierarchical models. With the availability of efficient and open source Monte-Carlo samplers no more hurdles exists to embark on an exploration of the full posterior distribution of such realistic hierarchical models.

I presented two concrete examples, where Bayesian inference helps us in furthering our understanding of strongly interacting matter. The first is the extraction of lattice QCD spectral functions, which is an ill-posed inverse unfolding problem similar to the unfolding of detector tracks. Here the main challenges reside in the design of improved prior distributions that incorporate analytic properties of the spectra known from QCD, where machine learning may play an important role in the near future. At the same time there exists the need to improve the information content of simulated correlators. The use of anisotropic lattices or exploring more unconventional strategies, such as extensions of the multilevel algorithm to full QCD or deformed temporal paths in the complex plane are currently being considered. The second example concerned the estimation of model parameters for the QGP from a systematic comparison with experimental measurements from HICs. The central focus there lay on the efficient sampling of the likelihood, which was achieved by use of a Gaussian process model.

This talk is only able to scratch at the surface of the many fascinating ways how (Bayesian) statistical analysis can help us push forward the boundaries of our knowledge of strongly interacting matter. The interested reader is cordially invited to peruse the many other contributions to this conference in the realm of statistical methods and machine learning located in the corresponding dedicated session.

\bibliographystyle{JHEP}
\bibliography{Bayes}

\end{document}